\documentclass[pra,twocolumn,amsmath,amssymb,superscriptaddress,floatfix]{revtex4}
\usepackage{graphicx}
\usepackage{latexsym}
\usepackage{amstext}
\usepackage{amsxtra}
\usepackage{color}
\usepackage{pdfsync}
\usepackage{float}

\definecolor{darkblue}{rgb}{0, 0, 0.8}
\usepackage[colorlinks=true, breaklinks=true, linkcolor=darkblue, citecolor=darkblue, urlcolor=darkblue]{hyperref}

\begin{document}

\title{Single-atom trapping in holographic 2D arrays of microtraps with arbitrary geometries}
\author{F. Nogrette, H. Labuhn, S. Ravets, D. Barredo, L. B\'eguin, A. Vernier, T. Lahaye, and A. Browaeys}
\affiliation{Laboratoire Charles Fabry, Institut d'Optique, CNRS, Univ Paris Sud,\\
2 avenue Augustin Fresnel, 91127 Palaiseau cedex, France }
\date{\today}

\begin{abstract}
We demonstrate single-atom trapping in two-dimensional arrays of microtraps with arbitrary geometries. We generate the arrays using a Spatial Light Modulator (SLM), with which we imprint an appropriate phase pattern on an optical dipole trap beam prior to focusing. We trap single $^{87}{\rm Rb}$ atoms in the sites of arrays containing up to $\sim100$ microtraps separated by distances as small as $3\;\mu$m, with complex structures such as triangular, honeycomb or kagome lattices. Using a closed-loop optimization of the uniformity of the trap depths ensures that all trapping sites are equivalent. This versatile system opens appealing applications in quantum information processing and quantum simulation, e.g. for simulating frustrated quantum magnetism using Rydberg atoms. 
\end{abstract}

\maketitle

\section{Introduction}

Optical trapping of cold atoms~\cite{Ovchinnikov1998} allows for a variety of applications, from the study of quantum gases~\cite{reviewQG} to the manipulation of single atoms~\cite{Schlosser2001}. Impressive achievements in the engineering of quantum systems have been obtained using relatively simple configurations of light fields, such as single-beam traps~\cite{WeitzCO2}, crossed optical dipole traps~\cite{Barrett2001}, microlens arrays~\cite{Dumke2002,Schlosser2011}, optical lattices~\cite{Bloch2005,Nelson2007}, or speckle fields~\cite{Billy2008}.

In the last few years, an interest in more advanced tailoring of optical potentials has arisen. Several technical approaches can be considered. A first solution consists in ``painting'' arbitrary patterns of light using a time-dependent light deflector~\cite{Fatemi2007,Henderson2009}, over timescales that are fast compared to the typical oscillation frequency in the trap. Ultracold atoms then experience an optical potential corresponding to the time-averaged light intensity. Another approach relies on the generation of reconfigurable light patterns using spatial light modulators (SLM), either in amplitude or in phase~\cite{Gaunt2012,Gaunt2013,Boyer2004,Bergamini2004}.

\begin{figure}[b]
\centering
\includegraphics[width=8.5cm]{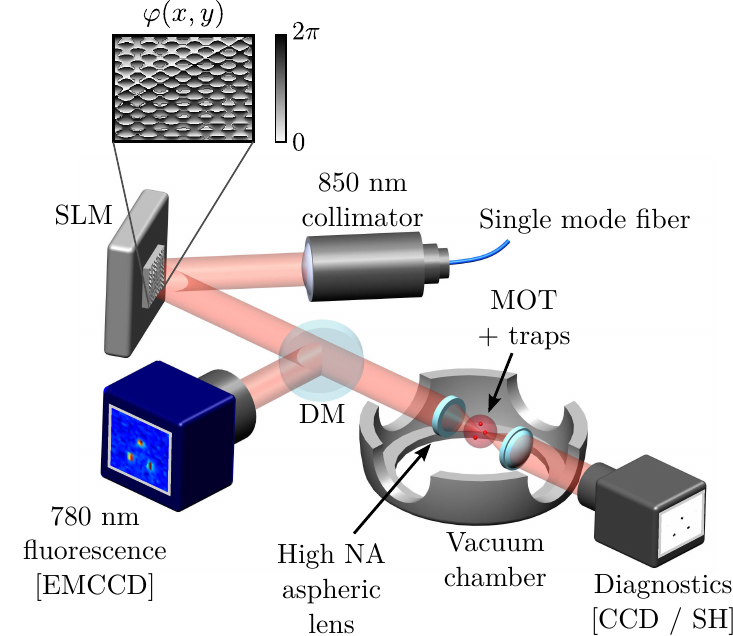}
\caption{Generation of an array of microtraps for single-atom trapping. The SLM imprints the calculated phase pattern $\varphi{(x,y)}$ on the 850~nm dipole trap beam. A high numerical aperture aspheric lens under vacuum focuses it at the center of a MOT. The intensity distribution in the focal plane is $\propto|{\rm FT}(A_0{\rm e}^{i\varphi})|^2$, where $A_0$ is the initial Gaussian amplitude profile of the 850~nm beam, and ${\rm FT}$ stands for Fourier Transform. The atomic fluorescence at 780~nm is reflected off a dichroic mirror (DM) and detected using an EMCCD camera. A second aspheric lens (identical to the first one) recollimates the 850~nm beam. This transmitted beam is used for trap diagnostics (either with a diagnostics CCD camera or a Shack-Hartmann (SH) wavefront sensor).}
\label{fig:fig1}
\end{figure}

Single atoms held in arrays of microtraps with a spacing of a few $\mu$m are a promising platform for quantum information processing and quantum simulation with Rydberg atoms~\cite{SaffmanRMP,Urban2009,Gaetan2009,Beguin2013,Barredo2014}. The realization of an array of $~\sim 50$ microtraps for single atoms using an elegant combination of fixed diffractive optical elements and polarization optics was recently demonstrated in~\cite{Piotrowicz2013}.

\begin{figure*}[t!]
\centering
\includegraphics[width=17cm]{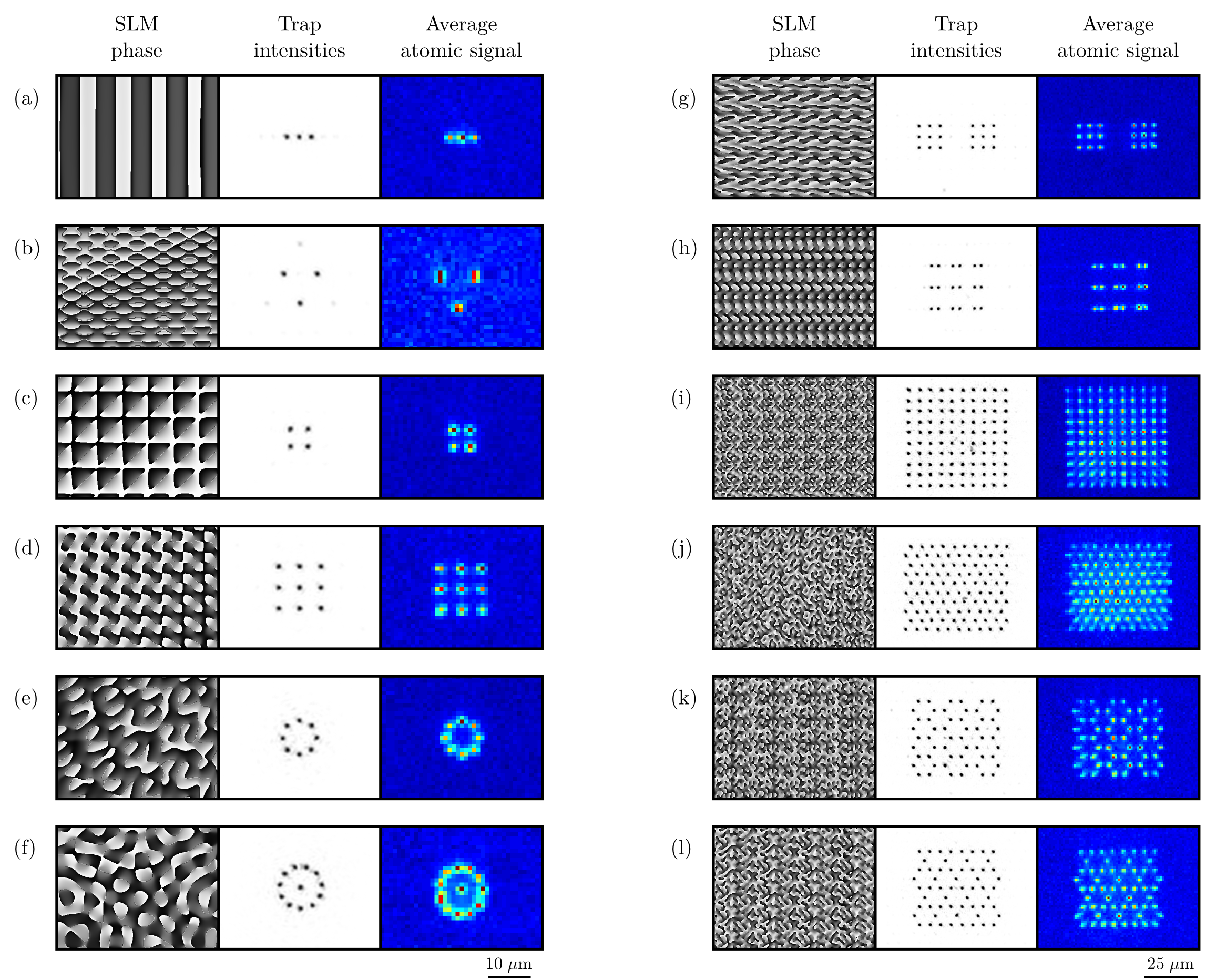}
\caption{A gallery of microtrap arrays with different geometries. For each panel, we show the calculated phase pattern~$\varphi$ used to create the array (left), an image of the resulting trap arrays taken with the diagnostics CCD (middle), and the average of $\sim 1000$ fluorescence images of single atoms loaded into the traps (right).}
\label{fig:imagesTrapGallery}
\end{figure*}

Here, we report on the trapping of single atoms in reconfigurable 2D arrays of  microtraps, separated by distances down to $3\;\mu$m, with almost arbitrary geometries. We create not only mesoscopic arrays of a few traps, but also regular 2D lattices with up to $\sim100$ sites, with geometries ranging from simple square or triangular lattices, to more advanced ones, such as kagome or honeycomb structures. Using a closed-loop optimization of the uniformity of the trap depths allows us to obtain very uniform lattices, which opens appealing prospects for quantum simulation with neutral atoms~\cite{blochdal} and eliminates a source of complication in the theoretical modeling of these systems. For that, we use a phase-modulating SLM, which has the advantage of being versatile and easily reconfigurable. A major asset of the system lies in the fact that, in combination with wavefront analysis, the SLM can also be used  to correct \emph{a posteriori} for aberrations that are inevitably present in the optical setup, thus improving considerably the optical quality of the traps. 

This article is organized as follows. After giving an overview of the principles behind our setup, we give a detailed account of the obtained results. We present a gallery of examples of microtrap arrays in which we trap single atoms, and we study the single-atom loading statistics of a $3\times3$ square array. In a second part, we give details about the implementation of the optical setup and the calculation of the phase holograms. We then explain how we optimize the obtained traps using a Shack-Hartmann (SH) wavefront sensor, and present a closed-loop improvement of the uniformity of the trap intensities. 

\section{Main results}

In this section, after briefly describing our experimental setup, we demonstrate the trapping of single atoms in microtrap arrays with various geometries.  

\subsection{Overview of the experimental setup}

Figure~\ref{fig:fig1} shows a sketch of the setup we use to trap single $^{87}$Rb atoms~\cite{Sortais2007}. It is based on a red-detuned dipole trap at a wavelength $\lambda=850$~nm, with a $1/e^2$ radius $w_{0}\simeq 1\,\mu$m. For a power of $3$~mW, the trap has a typical depth $U_{0}=k_{\rm B}\times 1$~mK, with radial (resp. longitudinal) trapping frequencies around 100~kHz (resp. 20~kHz). To load atoms into the microtrap, we produce a cloud of cold atoms at $\sim 50\:\mu{\rm K}$ in a magneto-optical trap (MOT). The dipole trap beam is focused in the cloud with a custom-made high-numerical aperture (NA) aspheric lens with focal length $f_{\rm{Asph.}}=\:10\,\rm{mm}$~\cite{lens}. We detect single atoms by measuring their fluorescence signal at 780~nm (collected by the same aspheric lens) using a cooled, 16--bit EMCCD camera~\cite{emccd}. We separate the fluorescence signal from the trapping beam with a dichroic mirror (DM). A second aspheric lens, facing the first one in a symmetrical configuration, is used to recollimate the trapping beam. An 8--bit CCD camera, placed after the vacuum chamber, is conjugated with the plane of the single atoms for diagnostic purposes.

We generate arrays of microtraps with arbitrary geometries using a phase-modulating SLM~\cite{slmnote}, which imprints a calculated phase pattern $\varphi(x,y)$ onto the trapping beam of initial Gaussian amplitude $A_0(x,y)$.  The intensity distribution in the focal plane of the aspheric lens is then given by the squared modulus of the 2D--Fourier transform of $A_0\exp\left( i \varphi \right)$. The phase pattern $\varphi$ needed to obtain a desired intensity distribution is determined by the iterative algorithm described in Sec.~\ref{subsec:HologramCalc}.

\subsection{Gallery of microtrap arrays}
\label{sec:Gallery}

Figure~\ref{fig:imagesTrapGallery} presents a selection of 2D trap arrays that we have created with the setup described above. For each array, we show the phase pattern $\varphi(x,y)$ used to create it, an image of the array obtained with the diagnostics CCD camera behind the chamber, and the average of $\sim 1000$ images of the atomic fluorescence  of single atoms in the traps (imaged with the  EMCCD camera). The figure illustrates strikingly the versatility of the setup. We can create small clusters containing $\sim 10$ traps, useful for the study of mesoscopic systems (a--h). It is also possible to create larger, regular lattices of up to $\sim100$~traps with varying degrees of complexity, from simple square (i) or triangular (j) lattices, to honeycomb (k) or kagome (l) structures, which opens for instance the possibility to simulate frustrated quantum magnetism with Rydberg-interacting atoms. The typical nearest-neighbor distance $a$ in those arrays is 4 to $5\;\mu{\rm m}$. We have also created arrays with spacings as small as $a\simeq3\;\mu{\rm m}$ without observing a significant degradation in the quality of the arrays. Other configurations, e.g. aperiodic structures, can be generated easily.

The total power needed to create an array of $N$  microtraps with a depth $U_0/k_{\rm B}\simeq 1\;{\rm mK}$ necessary for single-atom trapping is about $\sim 3N\,{\rm mW}$ on the atoms. Due to the finite diffraction efficiency of the SLM and losses on various optical components, we find that this requires to have slightly below $\sim 5N\,{\rm mW}$ at the output of the fiber guiding the 850~nm light to the experiment, which remains a very reasonable requirement even for $N=100$ traps.

\begin{figure}[t!]
\centering
\includegraphics[width=8.5cm]{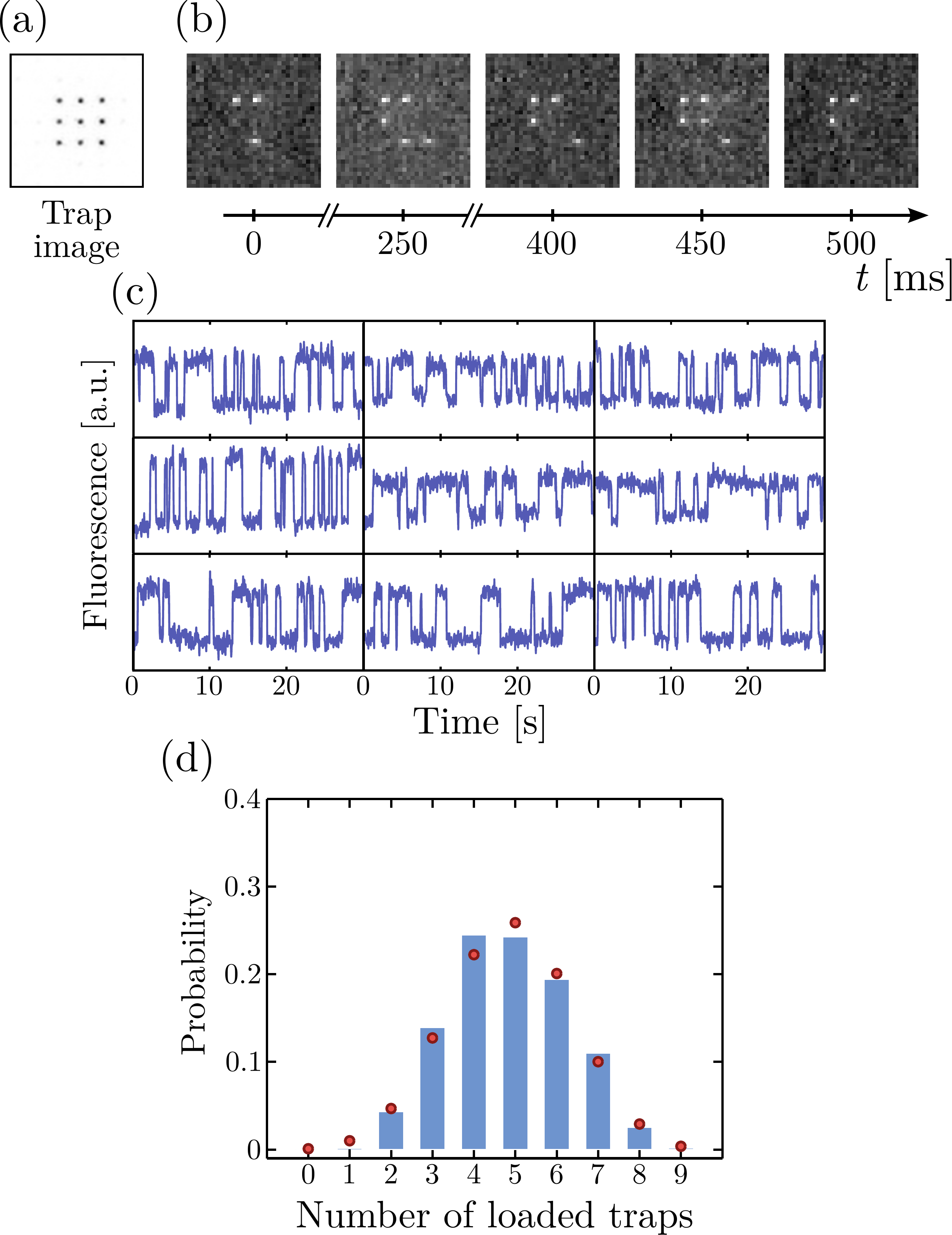}
\caption{Single atom trapping in a $3\times3$ array. (a) Image of the traps, separated by $4\;\mu$m, obtained with the diagnostics CCD camera. (b) Sample fluorescence images of single atoms trapped in the array. The exposure time is 50~ms. (c) Photons counts per 50~ms at the pixels corresponding to each of the nine trap positions, as a function of time. The telegraphic nature of the signal, with only two fluorescence levels, is the signature of single-atom trapping. (d) Histogram of the occurrences of images with $n$ atoms trapped (with $0\leqslant n\leqslant 9$) over a set of $\sim 2500$ images. The red dots correspond to the binomial distribution (\ref{eq:bin}) with $p=0.53$.}
\label{fig:singleAtoms}
\end{figure}

\subsection{Single-atom trapping in the arrays}

We now demonstrate directly single-atom trapping in a $3\times3$ square array [see Fig.~\ref{fig:singleAtoms}(a)]. Figure~\ref{fig:singleAtoms}(b) shows a series of snapshots obtained with the  EMCCD camera (the exposure time being 50~ms), showing fluorescence images of single atoms. As each of the $N=9$ traps has a probability $p\sim1/2$ of containing one atom, we observe that most images correspond to a sparsely loaded array, with an average number of atoms present close to $Np=9/2$, and fluctuations corresponding to atoms randomly entering and leaving each trap. To confirm that these images do correspond to single-atom trapping, we plot  the photon counts per 50~ms in the pixels corresponding to the positions of each of the nine traps as a function of time [see Fig~\ref{fig:singleAtoms}(c)]. One observes the characteristic `telegraphic signal', with only two  fluorescence levels, which is the hallmark of single atoms loaded into the microtraps by the collisional blockade mechanism~\cite{Schlosser2001,Sortais2007}. By analyzing each of the nine traces, we find that the occupation probability $p_i$ of each trap $i$ is close to $1/2$ (we find probabilities $p_i$ ranging from 0.43 to 0.57, with an average $\bar{p}=0.53$).

Figure \ref{fig:singleAtoms}(d) is a histogram of the number of atoms trapped in the $3\times3$ array, obtained by analyzing $\sim 2500$ images~\cite{Piotrowicz2013}. For an array of $N$ independent traps, if each trap has the same probability $p$ to be filled, the probability $P_n$ to have $n$ atoms in the array is given by the  binomial distribution
\begin{equation}
P_n=\frac{N!}{n!(N-n)!}p^n(1-p)^{N-n}.
\label{eq:bin}
\end{equation}
The dots on Fig.~\ref{fig:singleAtoms}(d) correspond to Eq.~(\ref{eq:bin}) with $N=9$ and $p=\bar{p}$ and show good agreement with the data. Therefore, the assumption that all traps are loaded with the same probability is a good approximation for estimating the probability of a given configuration to occur.

\section{Detailed implementation}

In the preceding section we focused on giving a detailed presentation of the results obtained. However, obtaining arrays of traps with as high a quality as what is demonstrated in Figs.~\ref{fig:imagesTrapGallery} and \ref{fig:singleAtoms} requires some care in the implementation of the setup. In this section, we detail the implementation of both the hardware and the software parts of the system. 

\begin{figure}[t]
\centering
\includegraphics[width=55mm]{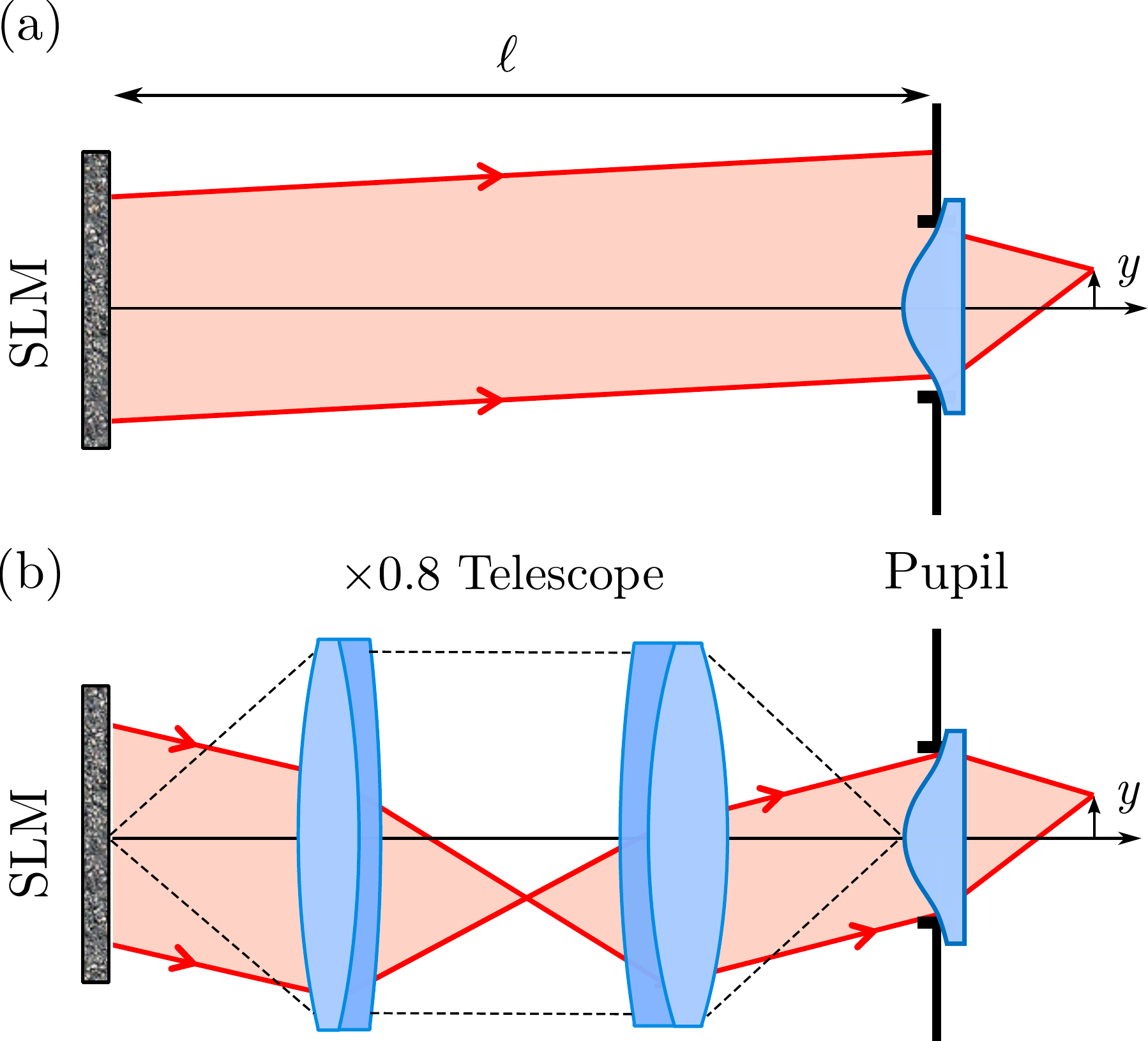}
\caption{Pupil conjugation. (a): Without a telescope, for a given field $y\neq 0$, the dipole trap beam is clipped and not centered on the aspheric lens. (b): The implemented telescope adapts the size of the beam to the aspheric lens pupil; by conjugating the SLM aperture to the entrance pupil of the aspheric lens, the beam is well centered, whatever the field.}
\label{fig:figConjPup}
\end{figure}

\subsection{Optical layout}

Our SLM has an active area of $12\times18\,\rm{mm^2}$, with a resolution of $600\times800$ pixels. It is illuminated by a collimated Gaussian beam with a $6.7$~mm $1/e^2$ radius coming from a polarization-maintaining, single-mode fiber connected to a collimator with a focal length $f=75$~mm. As diffraction-limited operation of the aspheric lens is obtained for an infinite-focus conjugation, with a pupil diameter~$D=10\:\rm{mm}$, we use an afocal telescope with a transverse magnification $m_{y}=-0.8$ to adapt the SLM active area to the aspheric lens aperture, while maintaining the collimation of the beam.

The implementation of the full system (vacuum chamber, dichroic mirror for fluorescence detection, components for generating the microtrap array) results in a relatively long distance ($\ell\simeq\:500~\rm{mm}$) between the SLM and the aspheric lens. This leads to the following problem (see Fig.~\ref{fig:figConjPup}(a)): when generating off-axis traps, the beam diffracted by the SLM impinges on the lens off-center, giving rise to clipping and field aberrations. This decreases the quality of arrays with a large number of microtraps. We circumvent this problem using \emph{pupil conjugation}: we take advantage of the extra degree of freedom given by the position of the telescope to conjugate the plane of the SLM with the aspheric lens, as shown in Fig.~\ref{fig:figConjPup}(b).

The optimization of the system is done with an optical design software. The simulation includes all the components from the optical fiber to the focal plane of the aspheric lens in the vacuum chamber. The lenses of the telescope and the lens of the collimator are near-infrared achromatic doublets used at low numerical aperture and small fields. The performance of the system over a field of $30\times30\;\mu{\rm m}^2$ in the microtrap plane is satisfactory: the Strehl ratio, i.e. the ratio of the actual peak intensity over the theoretical peak intensity for a diffraction-limited system~\cite{Gross2007}, is predicted to be ${\rm S}\geq 0.88 $ by the calculation.      

For the phase pattern calculation described below, we replace the telescope and the aspheric lens by a single equivalent lens with an effective focal length $f_{\rm eff}=f_{\rm Asph.}/|m_{y}|=12~\rm{mm}$ and an effective pupil in the SLM plane with diameter $D_{\rm eff}=12\:\rm{mm}$.

\subsection{Gerchberg-Saxton algorithm}
\label{subsec:HologramCalc}

\begin{figure}[t!]
\centering
\includegraphics[width=8.5cm]{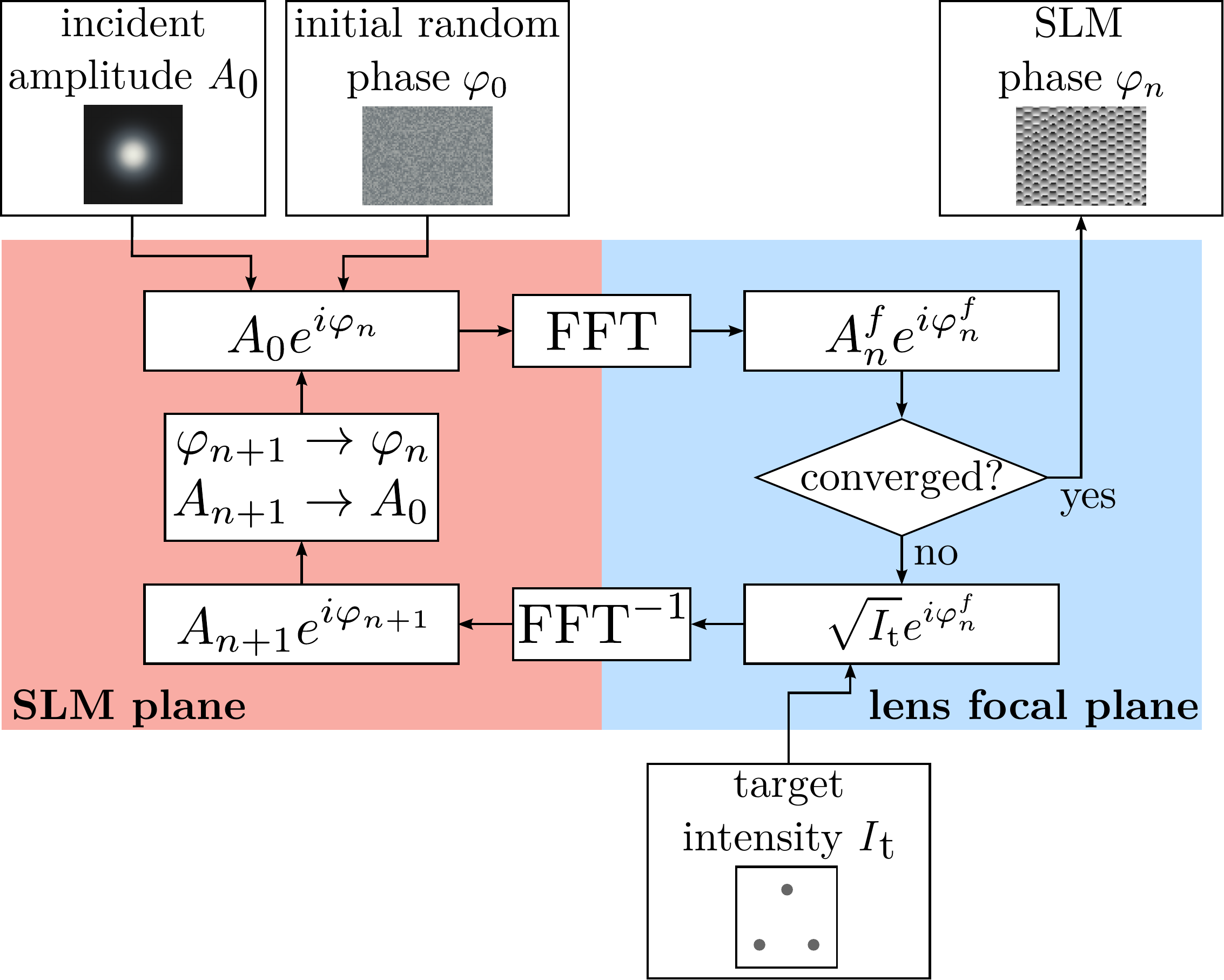}
\caption{The Gerchberg-Saxton algorithm. The field in the lens focal plane is calculated by Fast Fourier Transform (FFT) of the complex field in the SLM plane. If the obtained intensity $|A_{n}^{f}|^2$ does not match the target intensity $I_{\rm t}$, another iteration must be performed: the amplitude of the field in the focal plane is forced to the target amplitude $\sqrt{I_{\rm t}}$, and this new field is propagated back to the pupil plane by inverse FFT, resulting in a new amplitude and a new phase $\varphi_{n+1}$. This new phase is kept as the next SLM phase pattern, while the amplitude is forced to the incident one $A_{0}$, giving a new input field $A_{0}e^{i \varphi_{n+1}}$ for the next iteration.}
\label{fig:figGSAA}
\end{figure}

We use the Gerchberg--Saxton algorithm~\cite{Gerchberg1972} to calculate the phase pattern $\varphi(x,y)$ required to obtain an intensity distribution in the lens focal plane close to a desired target intensity $I_{\rm t}$. For the sake of completeness, we briefly recall below the essential steps of the algorithm (see Fig.~\ref{fig:figGSAA}).

We initialize the algorithm using a random phase pattern $\varphi_0$ in which each pixel value is given by a uniformly distributed random variate in the range $(0,0.2)\times 2\pi$. The target image $I_{\rm t}$ is a superposition of Gaussian peaks with $1/e^{2}$ radii $w=1\;\mu$m centered on the desired location of the microtraps. The amplitude of each Gaussian can be defined separately: this allows for correcting non-uniformities in the depths of the microtraps over the array (see section~\ref{sec:closed:loop}).

The incident field on the SLM is modeled as having a uniform phase and an amplitude $A_{0}(x,y)$. At each iteration of the algorithm, we propagate the electric field in the SLM plane $A_{0}e^{i\varphi_n}$ through the effective lens using Fast Fourier Transform (FFT) to calculate the field $A_n^{f}e^{i\varphi_n^f}$ in the focal plane. If the difference between the calculated intensity $|A_{n}^{f}|^2$ and the desired target image $I_{\rm t}$ is small enough, the phase pattern $\varphi_{n}$ is used to drive the SLM; otherwise, the amplitude of the field in the focal plane is replaced by the target amplitude $\sqrt{I_{\rm t}}$. This new field $\sqrt{I_{\rm t}} e^{i\varphi_n^f}$ is then propagated back to the SLM plane by inverse FFT, giving the field $A_{n+1}e^{i\varphi_{n+1}}$ in the SLM plane. The calculated phase $\varphi_{n+1}$ is kept as the new phase pattern in the SLM plane, while the amplitude is replaced by the incident one $A_0$, and another iteration is performed for the field $A_{0}e^{i \varphi_{n+1}}$. For the patterns shown in Fig.~\ref{fig:imagesTrapGallery}, the algorithm converges (i.e. the calculated phase patterns do not evolve any more) towards an approximate solution after typically a few tens of iterations~\cite{matlab}. The intensity distribution in the lens focal plane is then a good approximation of $I_{\rm t}$. However, we can approach the target even closer as described in section~\ref{sec:closed:loop}.

\subsection{Phase patterns displayed on the SLM}

\begin{figure}[t!]
\centering
\includegraphics[width=90mm]{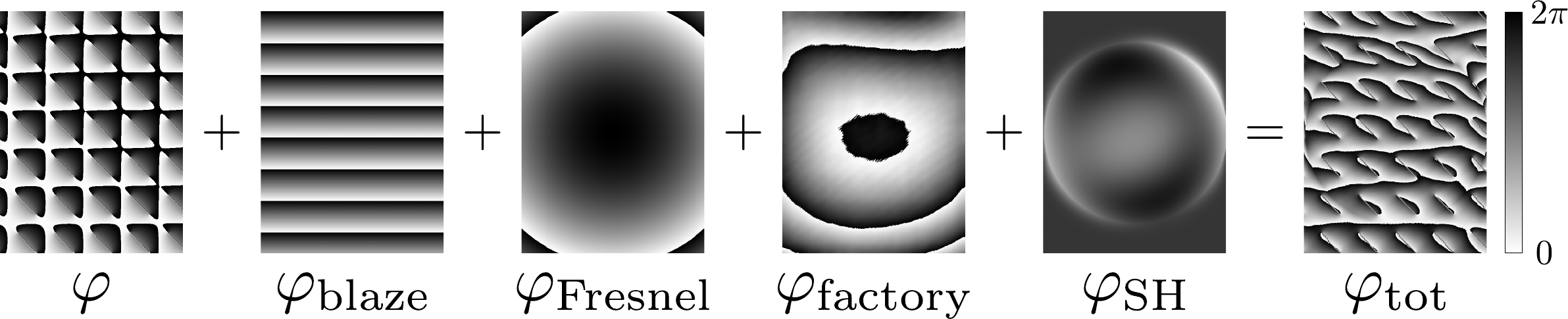}
\caption{Composition of the phase pattern $\varphi_{\rm tot}$ displayed on the SLM for generating the trap array of Fig.~\ref{fig:imagesTrapGallery}(c). The sum is calculated modulo $2\pi$.}
\label{fig:phase_pattern_composition}
\end{figure}

The phase pattern $\varphi_{\rm{tot}}$ used to drive the SLM includes several contributions beyond the calculated phase pattern $\varphi$, and reads:
\begin{equation}
\label{equationPhiCorrection}
\varphi_{\rm{tot}}= \varphi+\varphi_{\rm{blaze}} +\varphi_{\rm{Fresnel}}+ \varphi_{\rm{factory}}+\varphi_{\rm{SH}},
\end{equation}
where the sum is calculated modulo $2\pi$. In this equation,
\begin{itemize}
\item $\varphi_{\rm blaze}$ is a blazed grating pattern, allowing us to block the zeroth-order reflection from the SLM arising from its non-perfect diffraction efficiency;
\item $\varphi_{\rm Fresnel}$ is a quadratic phase pattern acting as a Fresnel lens, which allows us to fine-tune the focusing of the microtraps;
\item $\varphi_{\rm factory}$ is the correction phase pattern provided by the SLM manufacturer to correct for the optical flatness defects of the SLM chip;
\item $\varphi_{\rm SH}$ corrects for aberrations introduced by the setup and is obtained using a Shack-Hartmann wavefront sensor as described in section~\ref{sec:SHCorr} below.
\end{itemize}
Figure~\ref{fig:phase_pattern_composition} gives an example of the composition of the final phase pattern obtained by summing (modulo $2\pi$) the various terms described above.

\subsection{Improving the traps by analyzing the wavefront and correcting for aberrations using the SLM}
\label{sec:SHCorr}

Without the last term of Eq.~\eqref{equationPhiCorrection}, we observe that the quality of the obtained microtrap arrays decreases when the number of traps increases. Indeed, the assumption of a perfect effective lens used in the calculation of the hologram is not valid. The imperfections of the optics (vacuum windows, aspheric lens\ldots) and the residual misalignments deform the wavefront, thus reducing the depth of the microtraps.

\subsubsection{Wavefront measurement}
\label{subsec:SHMeasure}

\begin{figure}[t]
\centering
\includegraphics[width=6.1cm]{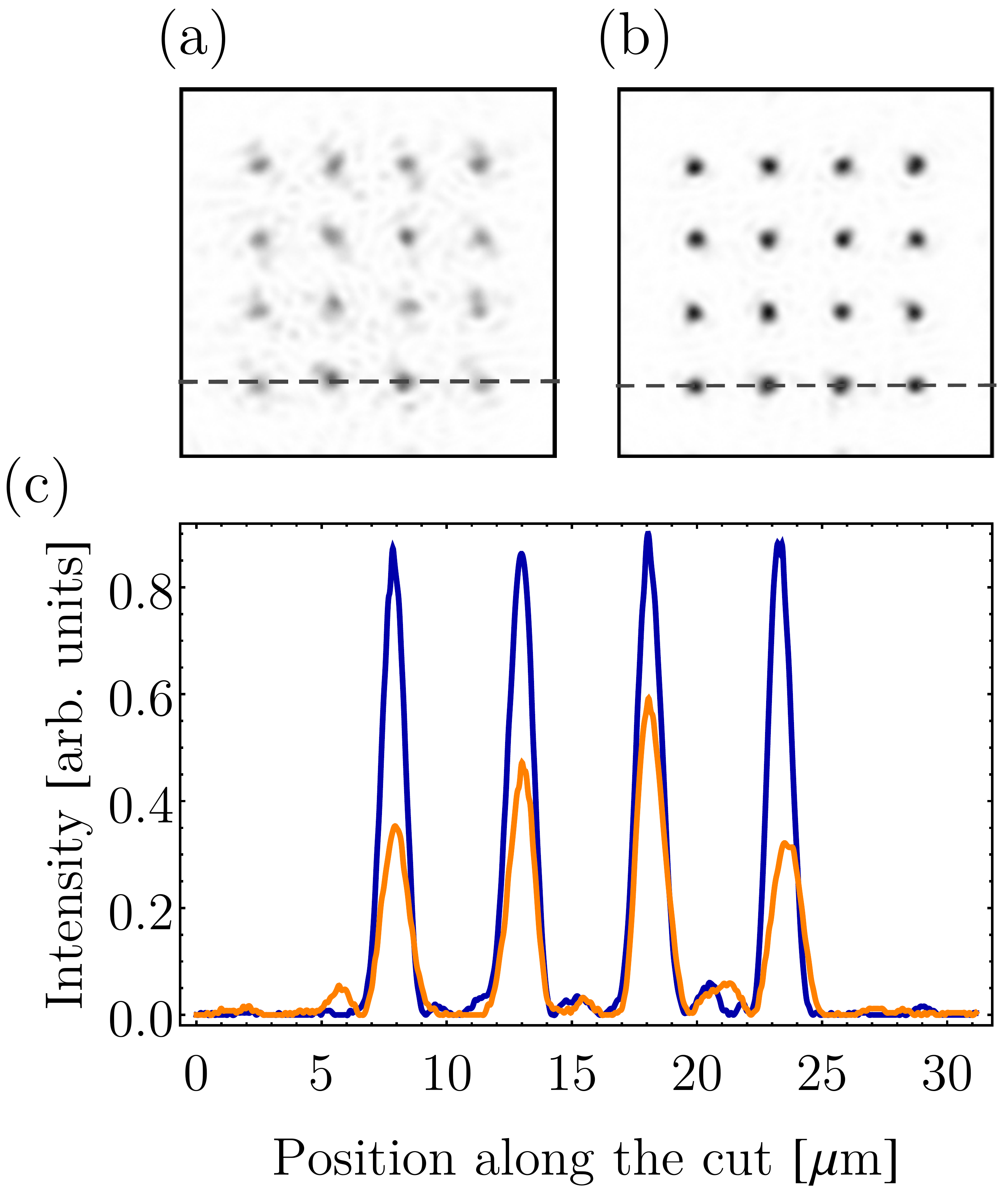}
\caption {Effect the Shack-Hartmann correction pattern $\varphi_{\rm SH}$. A CCD image of $4 \times 4$ microtraps is shown (a)~only with the factory correction and (b)~both the factory and the Shack-Hartmann patterns applied. (c): Intensity profiles along the dashed lines on (a--b), with (blue curve) and without (orange curve) correction $\varphi_{\rm SH}$. The arrays are created with the same calculated phase~$\varphi$. The laser power and the exposure time of the CCD camera are the same for both cases.}
\label{fig:correction}
\end{figure}

In order to correct for the above-mentioned imperfections, we measure the wavefront with a Shack-Hartmann sensor, and use the resulting $\varphi_{\rm SH}$ to drive the SLM~\cite{Lopez-Quesada2009}. We perform this measurement at the exit of the vacuum chamber, where the trapping beam has been recollimated by the second aspheric lens (see Fig.~\ref{fig:fig1}). The wavefront sensor~\cite{Haso} analyzes the wavefront corresponding to a single trap centered in the field where the phase pattern displayed on the SLM is $\varphi_{\rm blaze}+\varphi_{\rm factory}$. The measured rms deviation from a flat wavefront is $\delta_{\rm rms} = 0.15 \lambda$ (tilt and focus terms being removed). After applying the correction phase $\varphi_{\rm SH}$ to the SLM, we measure $\delta_{\rm rms} = 0.014 \lambda$. Figure~\ref{fig:correction} illustrates the impact of the phase corrections on the trap pattern (as measured by the diagnostics CCD camera) for a $4\times 4$ array: a comparison between panels (a) and (b) suggests that the correction increases the trap depth by a factor close to two.

This wavefront measurement includes the aberrations induced by the recollimating aspheric lens and the second vacuum window (see Fig.\ref{fig:fig1}). An independent wavefront measurement on the trapping beam before the chamber yields $\delta_{\mathrm{rms}}=0.05 \lambda$ without correction, showing that the optics of the vacuum chamber account for most of the wavefront aberrations. Applying directly the measured $\varphi_{\rm SH}$ on the SLM thus ``overcorrects'' aberrations, and one might fear that at the location of the atoms, the effect of the correction is actually detrimental. It is therefore desirable to check directly the actual effect of the correction on the atoms. For this purpose, we measure the trap depth and frequency directly with single atoms.

\subsubsection{Impact on the trap depth }
\label{subsec:SHLightShift}

\begin{figure}[t]
\centering
\includegraphics[width=6.6 cm]{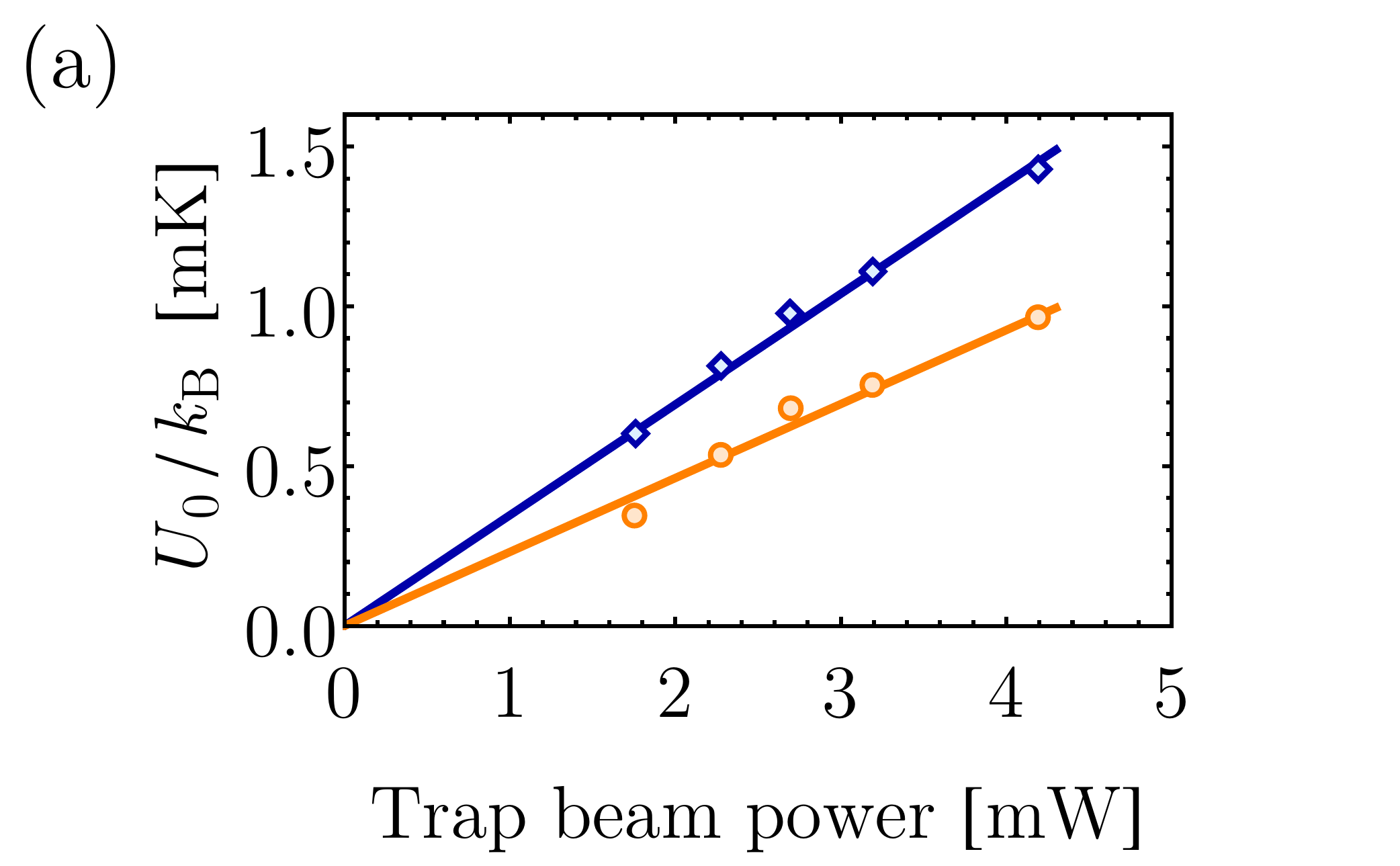}
\includegraphics[width=6.6 cm]{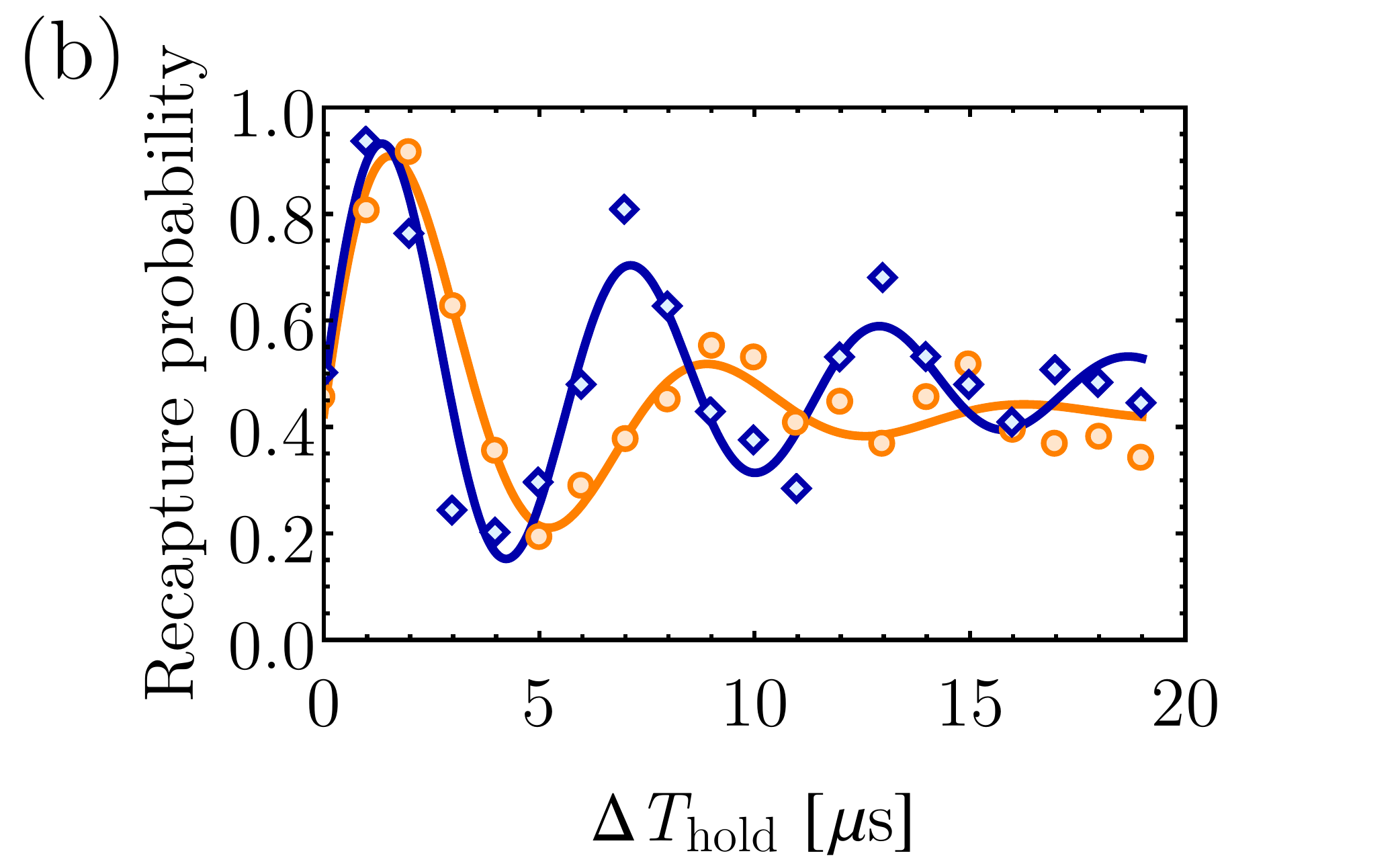}
\caption{ (a): Trap depth $U_0/k_{\rm B}$ as a function of the trap power, with (blue diamonds) and without (orange disks) Shack-Hartmann correction. With the latter, the trap depth increases by about $50\,\%$. (b): Recapture probabilities for an atom oscillating in the trap as a function of the hold time $\Delta T_{\mathrm{hold}}$. The trap frequency increases by about $30\,\%$ when the the Shack-Hartmann correction pattern is added to the SLM.}
\label{fig:lightshift-trapfreq}
\end{figure}

We measure the trap depth using light-shift spectroscopy with a single atom~\cite{Tey2008,Shih2013}. For that, we shine a $\sigma^+$ polarized probe that is quasi-resonant with the transition $|5S_{1/2},F=2,m_F=2 \rangle \to |5P_{3/2},F=3,m_F=3 \rangle$ on the atom and we record the number of fluorescence photons scattered by the atom as a function of probe detuning. The shift of the resonance with respect to its free-space value gives directly the trap depth $U_0$~\cite{Theselucas}. Figure~\ref{fig:lightshift-trapfreq}(a), obtained on the central trap of a $3\times 1$ array with a $4\;\mu$m separation, shows that including the Shack-Hartmann correction actually increases the trap depth by about $50\,\%$. 

\subsubsection{Impact on the trap frequency}
\label{subsec:SHTrapFreq}

Another important parameter of the trap is the trapping frequency. In order to determine the transverse trapping frequency seen by the atoms, we excite the breathing mode as in~\cite{Engler2000,Sortais2007}. For that purpose, the microtrap is switched off for a few microseconds, during which the atom leaves the center of the trap. When the trap is switched on again for a time $\Delta T_{\mathrm{hold}}$, the atom oscillates in the trap, with a radial frequency $\omega_{r}$~\cite{radial}. If the trap is then switched off again for a short time, the probability to recapture the atom afterwards depends on its kinetic energy at the time of the last switch-off, and thus oscillates at $2\omega_r$.

Figure~\ref{fig:lightshift-trapfreq}(b) shows the results of such a measurement, for a power of 2.8~mW per trap, again in the $3\times1$ array. The measured trap frequencies are $\omega_{r} = 2\pi \times 68.0$~kHz before correction, and $\omega_{r}=2 \pi \times {86.5}$~kHz with the Shack-Hartmann correction applied to the SLM. The increase in trapping frequency comes essentially from the increased depth of the corrected traps.

Using the single atom as a diagnostics tool, we could in principle test whether one can improve even further the trap quality by applying to the SLM a phase $\alpha \varphi_{\rm SH}$ (where $0\leqslant \alpha\leqslant 1$ is an adjustable parameter), in the hope of correcting only the aberrations `seen' by the atom, i.e. not the aberrations induced by the second lens and the second viewport. A test for $\alpha=1/2$ (which would yield the best correction if both lenses and windows introduced equal aberrations) gave results slightly worse than for $\alpha=1$, and in the following we thus keep this choice.  

\subsection{Closed-loop optimization of the uniformity of the trap depths in the array}
\label{sec:closed:loop}

\begin{figure}[t]
\centering
\includegraphics[width=70mm]{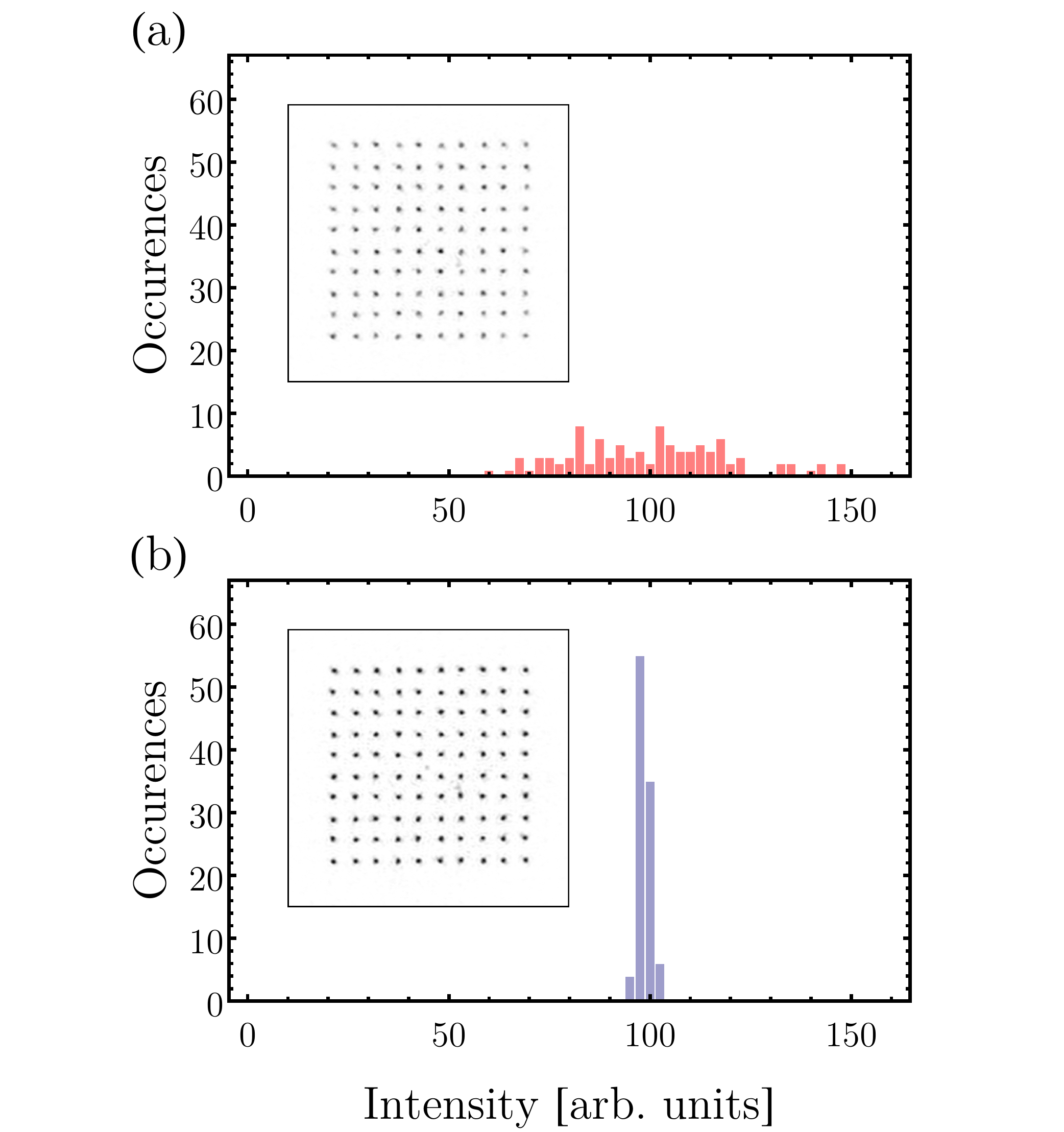}
\caption{Improving the uniformity of trap depths in a $10\times10$ square array. (a) Histogram of the maximal intensity levels of the microtraps $I_i$, measured with the diagnostics CCD camera (see inset), for the trap array obtained after a single use of the GS algorithm, and a target image where all traps have the same intensity. The standard deviation is 19~\%. (b) Same as (a) but after the closed-loop optimization of the uniformity of the trap intensities. The standard deviation is now 1.4~\%.}
\label{fig:histo}
\end{figure}

An important figure of merit to assess the quality of the arrays is the uniformity of the trap depths. Figure~\ref{fig:histo}(a) shows the distribution of the trap intensities, inferred from an analysis of an image of the array obtained with the diagnostics CCD camera, for a $10\times 10$ square lattice with a spacing $a=4\;\mu{\rm m}$. In this case, the phase applied to the SLM was obtained by running the GS algorithm with a target image $I_{\rm t}$ for which all traps have the same intensity. One observes a dispersion in the trap depths of $\pm\;19\%$ rms (the minimal and maximal values being $I_{\rm min}=61$ and $I_{\rm max}=148$, where the average intensity of all traps is normalized to $\bar{I}=100$). This variation in trap depths is detrimental for loading optimally the trap array with single atoms. Indeed, if the trap depth is too low, one still traps single atoms, but with a probability of occupancy significantly lower than $1/2$. Conversely, if the trap is too deep, one enters a regime in which the probability to have more than one atom is not negligible~\cite{Theselucas}. 

\begin{figure}[t]
\centering
\includegraphics[width=75mm]{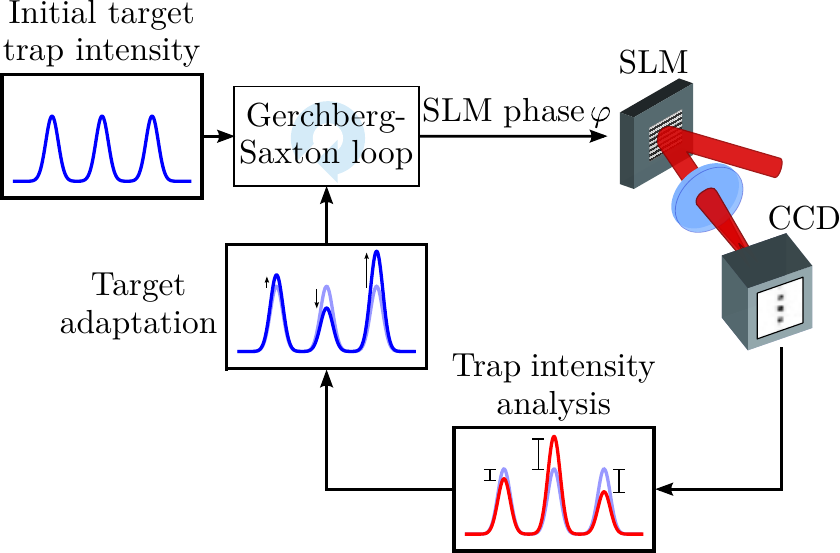}
\caption{Closed-loop algorithm used for improving the uniformity of trap depths. From the various trap intensities measured with the CCD camera (red profile) we calculate a new target intensity $I_{\rm t}$ following Eq.~(\ref{eq:target:adapt}): the brightest traps are dimmed, while the dimmest ones are enhanced. We then use this adapted target as the input for a new iteration of the GS algorithm, with the previously calculated phase as the initial condition.}
\label{fig:algo:unif}
\end{figure}

A way to compensate for this imperfection is to use the image of the trap array obtained with the diagnostics CCD to calculate a new target image where the new trap intensity $I'_i$ of trap $i$ is scaled according to the measured one $I_i$ as 
\begin{equation}
I'_i=\frac{\bar{I}}{1-G(1-I_i/\bar{I})},
\label{eq:target:adapt}
\end{equation}
where $\bar{I}$ is the average intensity of all traps and $G$ an adjustable ``gain''. In other words, traps that are two weak get enhanced in the new target image, while the brightest ones get dimmed. We then run again the GS algorithm with this new target image as an input, and with the previously obtained phase pattern $\varphi$ as the initial guess for the phase (see Fig.~\ref{fig:algo:unif}). We observe that the distribution of the trap intensities decreases quite drastically after a few iterations. Choosing $G\simeq 0.7$ gives the best performance (lower values decrease the convergence speed, while higher values yield to overshoots in the correction). Figure~\ref{fig:histo}(b) shows the resulting histogram of trap intensities for the $10\times10$ square lattice, after 20 iterations. The array is now very uniform, with trap intensities varying between 96~\% and 103~\% of $\bar{I}$ (peak-to-peak). This corresponds to a 15-fold reduction in the dispersion of the trap depths. 

The single-atom trapping demonstrated in the arrays of Figs.~\ref{fig:imagesTrapGallery} and \ref{fig:singleAtoms} could be achieved only after this closed-loop optimization was implemented, and illustrates strikingly the efficiency of the method. We believe that such an optimization, which takes full advantage of the reconfigurable character of the SLM, could prove useful in order to create very uniform lattices with arbitrary structures for quantum simulation with ultracold atoms.

\section{Conclusion and outlook}

The simple setup described above is a versatile tool for creating arrays of microtraps with almost arbitrary geometries. We have demonstrated single-atom loading in such arrays, which opens exciting possibilities to engineer interesting few-atom entangled states using e.g. Rydberg blockade~\cite{MuellerGate}, especially in combination with dynamical addressability using moving optical tweezers~\cite{Beugnon2007}.

For arrays with a large number of traps, a current limitation of the system is the non-deterministic character of the single-atom loading of the micro-traps: as each trap has a probability $1/2$ of being filled with an atom, a $N$-trap array has, at any given time, only an exponentially small probability $1/2^N$ to be fully loaded. Implementing quasi-deterministic loading schemes will thus be needed to take full advantage of the setup. Using Rydberg blockade, loading probabilities of $\sim 60\,\%$ have been recently demonstrated in a single microtrap~\cite{saffmanarxiv}. Alternatively, using a blue-detuned `collision beam', relatively high loading probabilities, in excess of $80\,\%$, have been achieved~\cite{andersen}, which opens the way to loading arrays of a few tens of traps over reasonable timescales.  

In combination with the recently demonstrated Raman sideband cooling of single atoms trapped in optical tweezers~\cite{regal,lukin}, a similar system with smaller distances between microtraps ---which could be achieved using high-numerical apertures objectives such as the ones used in quantum gas microscopes~\cite{greiner}--- could then become an interesting alternative approach to study the many-body physics of ultracold atoms in engineered optical potentials, without using traditional optical lattices~\cite{regalarxiv}.

\begin{acknowledgments}
We thank Yvan Sortais for invaluable advice about the optical design and for a careful reading of the manuscript, Andr\'{e} Guilbaud for technical assistance, and Bruno Viaris, Laurence Pruvost and Zoran Hadzibabic for fruitful discussions. We are grateful to Lionel Jacubowiez, Thierry Avignon, and Samuel Bucourt for the loan of Imagine Optic Shack-Hartmann wavefront sensors. This work was supported financially by the EU (ERC Stg Grant ARENA, AQUTE Integrating project, FET-Open Xtrack project HAIRS, EU Marie-Curie program ITN COHERENCE FP7-PEOPLE-2010-ITN-265031 (H.L.)), by the DGA (L.~B.), and by R\'egion \^Ile-de-France (LUMAT and Triangle de la Physique, LAGON project). 
\end{acknowledgments}


\begin{thebibliography}{99}
\bibitem{lens}
We use a D-ZLaF52LA aspheric lens from LightPath technologies. It is optimized at 780~nm for an infinite-to-focus conjugation with a numerical aperture ${\rm NA}=0.54$. When used at 850~nm, its performance is maintained for a reduced aperture ${\rm NA}=0.44$, i.e. $D=10\,\rm{mm}$.
\bibitem{emccd}
We use a Andor iXon Ultra, which can be used in an EMCCD (Electron Multiplying CCD) mode. However, due to the relatively high photon yield ($\sim 300$~photons per $50$~ms time bin), here we do not need to use the electron multiplier to detect single atoms.
\bibitem{matlab}
We have implemented the algorithm using \textsc{Matlab}. Without any attempt to optimize the code, convergence takes less than one minute on a personal computer.
\bibitem{radial}
The longitudinal confinement is much weaker than in the radial direction, so for short switch-off times the longitudinal motion of the atom is only weakly excited.
\end{thebibliography}
\end{document}